\documentclass[aps,pra,twocolumn,showpacs,10pt]{revtex4-1}
\usepackage{graphicx}
\usepackage{amssymb}
\usepackage{amsmath}
\begin{document}
\title{Efficiency of light-frequency conversion in an atomic ensemble}
\author{H. H. Jen and T. A. B. Kennedy}
\affiliation{School of Physics, Georgia Institute of Technology, Atlanta, Georgia 30332, USA}
\date{\today}

\begin{abstract}The efficiency of frequency up and down conversion of light in an atomic ensemble, with a diamond level configuration, is analyzed theoretically. The conditions of pump field intensities and detunings required to maximize the conversion as a function of optical thickness of the ensemble are determined. The influence of the probe pulse duration on the conversion efficiency is investigated by numeric
solution of the Maxwell-Bloch equations.
\end{abstract}
\maketitle
\section{Introduction}
The frequency conversion of light fields has been an important theme in optical physics for around half a century.  In quantum information physics the
conversion of single photons to and from the telecom wavelength band is a topic of more recent vintage, and is motivated by the desire to minimize optical fiber transmission losses when distributing entangled states over distant quantum memory elements in a quantum repeater \cite{repeater}.

An associated technical problem is that telecom light is not readily stored in ground level atomic memory coherences. Retrieval processes in atomic
ensembles , for example using electromagnetically induced transparency \cite{eit}, or more specifically the dark-polariton mechanism \cite{Lukin,polariton}, generate shorter wavelength radiation correlated to the stored atomic excitation by Raman scattering. Such radiation, optically resonant to the ground level of typical atoms and ions, has been retrieved in numerous experiments \cite{qubit,chou,entangle,vuletic,retrieval,single,pan,kimble,ionent,ion}. An important advance would involve generation of atomic memory coherences quantum-correlated with telecom wavelength radiation, thereby minimizing transmission losses over long distances. Recently there has been a breakthrough in this direction using a pair of cold, nondegenerate rubidium gas samples \cite{radaev}. The stored excitation is correlated with an infrared field (idler) in one gas sample, and the idler is then frequency converted to a telecom wavelength signal field in the other ensemble. The frequency conversion mechanism involves the diamond configuration of atomic levels shown in Fig. 1.

In a probabilistic protocol it is important to maximize all efficiencies, e.g., fiber transmission, single-photon detection and quantum memory lifetime \cite{ran}. In the present work we investigate the efficiency of frequency up- and down- conversion in the diamond atomic configuration \cite{telecom,orozco}, as a function of the ensemble's optical thickness, and the intensity and detuning of the pump fields involved in the near-resonant, four-wave mixing process.

The remainder of this paper is organized as follows. In Sec. II we discuss the four wave mixing process and present solutions for the up- and down-
converted fields; the dressed state picture is used as a guide to understand the characteristic features of the absorption and signal-idler field coupling, as a function of the microscopic interaction parameters. In Sec. III we present results of an optimization of the conversion efficiency as a function of the optical depth of the atomic ensemble. In Sec. IV we investigate the effects of finite pulse duration by integrating the Maxwell-Bloch equations for the system. Section V presents our conclusions. The derivations of the Maxwell-Bloch and parametric equations are relegated to the Appendix.%

\begin{figure}[b]
\begin{center}
\includegraphics[width=9cm,height=6.5cm]{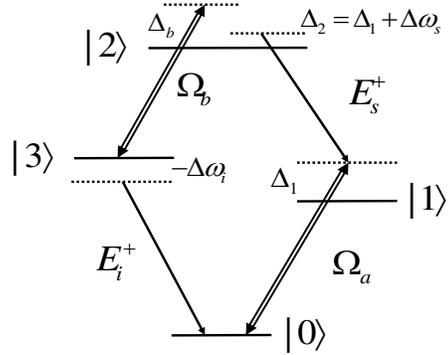}
\caption{The diamond configuration of atomic system for conversion scheme. Two pump lasers (double line) with Rabi frequencies $\Omega_a,\Omega_b$ and propagated probe fields (single line) $E_s^{+},E_i^{+}$ interact with the atomic medium. Various detunings are defined in the Appendix and the atomic levels used in the experiment \cite{radaev} are $(|0\rangle,|1\rangle,|2\rangle,|3\rangle)=(|5\text{S}_{1/2},\text{F}=1\rangle,|5\text{P}_{3/2},\text{F}=2\rangle,|6\text{S}_{1/2},\text{F}=1\rangle,|5\text{P}_{1/2},\text{F}=2\rangle).$}
\label{four}
\end{center}
\end{figure}

\section{Up and down conversion efficiency}

In this paper, we consider a cold and cigar-shaped $^{87}$Rb atomic ensemble with co-propagating light fields similar to the experimental setup in Ref.
\cite{radaev}.

The conversion scheme shown in Fig. \ref{four} involves two pump lasers with frequencies $\omega_{a}$ and $\omega_{b}$, respectively; their Rabi
frequencies are given by $\Omega_{a}$ and $\Omega_{b}$. Two weak probe fields, signal and idler, with frequency $\omega_{s}$ and $\omega_{i}$,
respectively, propagate through the optically thick atomic medium. Unlike the cascade driving scheme, where two-photon excitation generates a photon
pair spontaneously \cite{telecom}, pump laser b experiences a transparent medium if both the signal and idler fields are in the vacuum state. With an
incident signal field, four wave mixing with the pumps generates an up-converted idler field, while an incident idler field generates a down-converted signal.

The Maxwell-Bloch equations for the interacting system of light and four light fields is derived in the Appendix. By linearizing the equations with respect to the signal and idler field amplitudes, and adiabatically eliminating the atoms, one arrives at coupled parametric equations for the signal and idler fields. We discuss their solution in this section, and leave numerical solutions of the Maxwell Bloch equations to Sec. IV. The calculation of conversion efficiencies can also be carried out with the quantized Heisenberg-Langevin version of the coupled parametric equations. The resulting conversion efficiencies are identical to the semi-classical treatment; the additional quantum noise contributions vanish as the $|2\rangle\leftrightarrow |3\rangle$ transition driven by pump laser b has vanishing populations and atomic coherence. A similar simplification occurs in the calculation of the storage efficiency of spin waves in a system of atoms in the $\Lambda$ configuration \cite{gorshkov}.

The co-moving propagation equation for c-number signal and idler fields (respectively, ${E}_{s}^{+}$ and ${E}_{i}^{+}$) under energy conservation
($\Delta\omega=\omega_{a}+\omega_{s}-\omega_{b}-\omega_{i}=0$) and four-wave mixing conditions ($\Delta k=k_{a}-k_{s}+k_{b}-k_{i}=0$) are%
\begin{align}
\frac{d}{dz}{E}_{s}^{+}  &  =\beta_{s}{E}_{s}^{+}+\kappa_{s}{E}_{i}^{+},\nonumber\\
\frac{d}{dz}{E}_{i}^{+}  &  =\kappa_{i}{E}_{s}^{+}+\alpha_{i} {E}_{i}^{+}.
\end{align}
The coupled equations are similar to those found for the double $\Lambda$ system \cite{braje,couple}. The self-coupling coefficients $\beta_{s},$
$\alpha_{i}$ and parametric coefficients $\kappa_{s}$, $\kappa_{i}$ are defined in the Appendix. The solution, under conditions of down conversion with
boundary condition $E_{s}^{+}(0)=0$, is

\begin{align}
&
\begin{bmatrix}
E_{s}^{+}(L)\\E_{i}^{+}(L)
\end{bmatrix}
=\nonumber\\
&\frac{E_{i}^{+}(0)e^{(\alpha_{i}+\beta_{s})L/2}}{2w}%
\begin{bmatrix}
\kappa_{s}(e^{wL}-e^{-wL})\\
\frac{1}{(w+q)}[\kappa_{s}\kappa_{i}e^{wL}+(q+w)^{2}e^{-wL}]
\end{bmatrix},
\nonumber\\
&
\end{align}
where $w\equiv\sqrt{q^{2}+\kappa_{s}\kappa_{i}}$, and $q\equiv(-\alpha_{i}+\beta_{s})/2$. We define the down conversion efficiency $\eta_{\text{d}}$
and transmission of input idler field $T_{\text{d}}$ by

\begin{align}
\eta_{\text{d}}  &  =\left\vert \frac{E_{s}^{+}(L)}{E_{i}^{+}(0)}\right\vert
^{2}=\left\vert \frac{\kappa_{s}}{2w}e^{(\alpha_{i}+\beta_{s})L/2}(e^{wL}-e^{-wL})\right\vert ^{2}\label{down},\\
T_{\text{d}}  &  =\left\vert \frac{E_{i}^{+}(L)}{E_{i}^{+}(0)}\right\vert^{2}\nonumber\\
&  =\left\vert \frac{e^{(\alpha_{i}+\beta_{s})L/2}}{2w(w+q)}[\kappa_{s}\kappa_{i}e^{wL}+(q+w)^{2}e^{-wL}]\right\vert ^{2}.
\end{align}
For up-conversion, ($E_{i}^{+}(0)=0$) symmetry gives the corresponding coefficients

\begin{align}
\eta_{\text{u}}  &  =\left\vert \frac{E_{i}^{+}(L)}{E_{s}^{+}(0)}\right\vert
^{2}=\left\vert \frac{\kappa_{i}}{2w}e^{(\alpha_{i}+\beta_{s})L/2}(e^{wL}-e^{-wL})\right\vert ^{2}\label{up},\\
T_{\text{u}}  &  =\left\vert \frac{E_{s}^{+}(L)}{E_{s}^{+}(0)}\right\vert
^{2}\nonumber\\
&  =\left\vert \frac{e^{(\alpha_{i}+\beta_{s})L/2}}{2w(w+q)}[(q+w)^{2}e^{wL}+\kappa_{s}\kappa_{i}e^{-wL}]\right\vert ^{2}.
\end{align}

The up and down conversion efficiencies differ only by the interchange $\kappa_{i} \leftrightarrow\kappa_{s}.$ In the strong parametric coupling
regime where $|\kappa_{i}|,|\kappa_{s}| \gg|\alpha_{i}|,|\beta_{s}|$, the coefficients can be simplified to $\eta_{\text{u}} \simeq\sqrt{\frac
{\kappa_{i}}{\kappa_{s}}}\sinh(\sqrt{\kappa_{s}\kappa_{i}}L),\eta_{\text{d}}\simeq\sqrt{\frac{\kappa_{s}}{\kappa_{i}}}
\sinh(\sqrt{\kappa_{s}\kappa_{i}}L)$ and $T_{\text{u}}=T_{\text{d}} \simeq\cosh(\sqrt{\kappa_{s}\kappa_{i}}L).$ \

\begin{figure}[b]
\begin{center}
\includegraphics[width=10cm,height=7.5cm]
{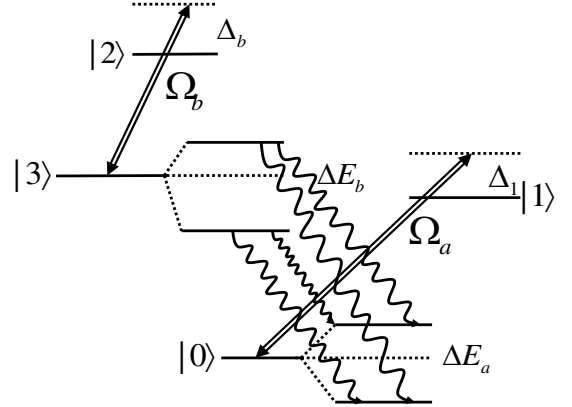}
\caption{Dressed-state picture from the perspective of the probe idler transition between atomic levels $|0\rangle$ and $|3\rangle.$ \ Two strong fields $\Omega_a,\Omega_b$ shift the levels with energy $\Delta E_{a,b} $ and wavy lines represent the idler field resonances.}
\label{dress}%
\end{center}
\end{figure}

Under the further assumptions $\alpha_{i}=\beta_{s}=0$ and $\kappa_{i},\kappa_{s}$ are pure imaginary, we find
$\eta_{\text{u}}=\eta_{\text{d}}$ $=\sin^{2}[\operatorname{Im}(\kappa_{s}L)]$ and
$T_{\text{u}}=T_{\text{d}}=\cos^{2}[\operatorname{Im}(\kappa_{s}L)]$. This result was recently derived by Gogyan using a dressed-state approach \cite{gogyan08}, in the case of resonant pump fields $\Delta_1= \Delta_b=0$ \cite{gogyan}. In this ideal limit there is a
conservation condition $\eta_{\text{u}}+T_{\text{u}}=\eta _{\text{d}}+T_{\text{d}}=1.$ \ The parametric coupling coefficients are
not identical, but in the regime of strong coupling they approach each other. As noted by Gogyan, when the pump-a intensity is
large ($\Omega_{a} >> |\Delta_{1}|, \gamma_{03}$) the $|0\rangle\leftrightarrow|1\rangle$ is
saturated, the atomic coherence is negligible and $\kappa_{s}\approx\kappa_{i}\propto\tilde{\sigma}_{00,s}(\frac{\Omega_{a}^{\ast}\Omega_{b}}{T_{02}%
}+\frac{\Omega_{a}^{\ast}\Omega_{b}}{T_{13}}) $. Alternatively, in the limit $\Omega_{b} >> \Omega_{a}, \gamma_{32}$ and
$|\Delta_{1}| >> \gamma_{01}$ the atomic coherence of $|0\rangle\leftrightarrow|1\rangle$ dominates and once again
$\kappa_{s}\approx\kappa_{i}\propto\frac{i\Omega _{b}|\Omega_{b}|^{2}\tilde{\sigma}_{01,s}^{\dag}}{T_{13}T_{02}}$.

The ac-Stark splitting induced by the pump lasers shifts the resonant absorption condition for the idler and signal fields. The idler and signal
experience resonant absorption at the transition frequency of the dressed atom. The corresponding transitions for the idler are shown in Fig.\ref{dress}. The bare states are shifted by $\Delta E_{a} =\left\vert \Delta_{1}\pm\sqrt{\Delta_{1}^{2}+4\Omega_{a}^{2}}\right\vert /2 $ and $\Delta E_{b}=\left\vert \Delta_{b}\pm\sqrt{\Delta_{b}^{2}+4\Omega_{b}^{2}}\right\vert /2, $ respectively. Note that our Rabi frequencies are smaller by a factor 2 than the standard definitions to avoid a plethora of prefactors in the equations of the Appendix.

For resonant pump fields, $\Delta E_{a,b}=\pm\Omega_{a,b}$ . The idler transition resonances are at
$\Delta\omega_{i}=-(\Omega_{a}+\Omega_{b}),$ $-\left\vert \Omega_{a}-\Omega_{b}\right\vert ,$ $\left\vert
\Omega_{a}-\Omega_{b}\right\vert ,$ $(\Omega_{a}+\Omega_{b})$ and these delineate three windows separated by these four
absorption peaks. For $\Omega_{a}>\Omega_{b}$ the centers of these windows are at $-\Omega_{a},$ $0,$ and $\Omega_{a}$,
respectively. Choosing the idler detuning $\Delta\omega_{i}=\pm\Omega_{a}$ as in \cite{gogyan}, the idler interacts with the
atomic medium at the center of the left or right window.
\begin{figure}[b]
\begin{center}
\includegraphics[width=7.5cm,height=8cm,angle=-90]{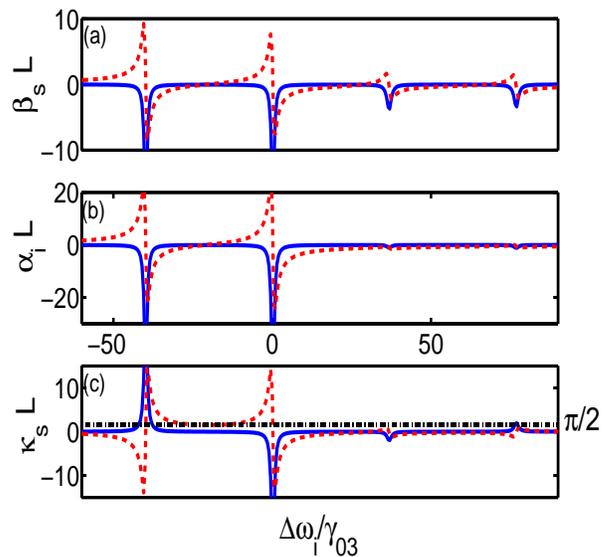}
\caption{(Color online) Self-coupling coefficients $\beta_{s},\alpha_{i}$ and cross-coupling coefficient $\kappa_{s}$. Dimensionless quantities (a) $\beta_{s}L$, (b) $\alpha_{i}L$ and (c) $\kappa_{s}L$ with real (solid blue) and imaginary (dashed red) parts are plotted as a dependence of idler detuning $\Delta\omega_{i}$ [same label in (b)] showing four absorption peaks to construct three parametric coupling windows. A black dashed-dot line of the constant $\pi/2$ is added in (c) to demonstrate the crossover with Im$(\kappa_{s}L)$ indicating the ideal conversion efficiency condition in the left window. The parameters we use are ($\Omega_{a}$, $\Omega_{b}$, $\Delta_{1}$, $\Delta_{b}$, $\Delta\omega_{i}$) $=(33$, $20$, $39$, $2$, $-21)\gamma_{03}$ for optical depth $\rho\sigma L=150$ with $L=6$mm. Various natural decay rates are $\gamma_{03}=1/27.7\text{ns}$, $\gamma_{01}=1/26.24\text{ns}$, $\gamma_{12}=\gamma_{03}/2.76$ and $\gamma_{32}=\gamma_{03}/5.38$ \cite{gsgi}.}
\label{coefficient}
\end{center}
\end{figure}

As an example of the strong coupling windows created by intense pump lasers, we show in Fig. \ref{coefficient} the self- and
 cross-coupling coefficients for the signal and idler fields as a function of the idler frequency. Note that the corresponding
 frequency of signal field is determined by $\Delta\omega_{s}=\Delta\omega_{i}-\Delta_{1}+\Delta_{b}$. \ The dimensionless
 quantities $\alpha_{i}L$, $\beta_{s}L$, and $\kappa_{s} L$ are shown under the conditions of maximum conversion efficiency
  to be discussed in the next section. \ We choose the optical depth (opd) $\rho\sigma L=150$ where $\rho$ is the number
   density, $\sigma\equiv3\lambda^{2}/(4\pi)$ the resonant absorption cross section, and $L$ the atomic ensemble length in the propagation
    direction. \ Three parametric coupling windows are separated by two strong absorption peaks on the left
    and two relatively weak ones on the right. The imaginary part of the self-coupling coefficients are seen to vanish in
    each window at a certain point, while the real parts are small away from resonances. At the same time the cross-coupling
    coefficients have a large imaginary part.\ The positive gradient of Im$(\beta_{s}L)$ and Im$(\alpha_{i}L)$, inside
    the windows is indicative of normal dispersion. \

\section{Optimal conversion efficiency}

It is important to ascertain the parameters which allow maximum efficiency of conversion due its potential in practical quantum
information processing. \ In principle we need to search the three parametric coupling windows to find the optimum
conditions for an atomic ensemble of a given optical thickness.

\begin{figure}[b]
\begin{center}
\includegraphics[width=5.0cm,height=8cm,angle=-90]{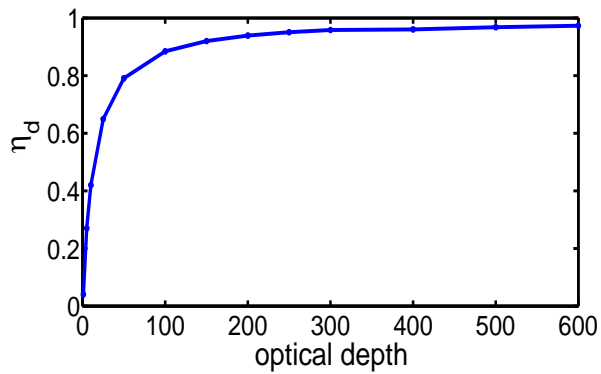}
\caption{(Color online) Down conversion efficiency $\eta_{\text{d}}$ vs optical depth (opd) from $1$ to $300.$ \ Each dotted point is the maximum for five variational parameters $\Omega_{a}$,$\Omega_{b}$, $\Delta_{1}$, $\Delta_{b}$, and $\Delta\omega_{i}$.}
\label{opd}
\end{center}
\end{figure}

In the previous section we have discussed how three parametric coupling windows appear for some particular values of pump laser parameters.\ In the search for the maximal conversion efficiency, five parameters $\Omega_{a}$, $\Omega_{b}$, $\Delta_{1}$, $\Delta_{b}$, and
$\Delta\omega_{i}$ are varied to maximize the conversion efficiency for a fixed optical depth of atomic ensemble, using functional optimization.

The optical depth $\rho\sigma L$ appears through the dependence on atomic number $N$ in the Arecchi-Courtens cooperation time $T_{c}$ \cite{scale}
\[
T_{c}^{-2} \equiv N|g_{i}|^{2} = \frac{\gamma_{03}c}{2 L} \ \ \rho\sigma L.
\]
\ In Fig. \ref{opd}, we show the maximum of down conversion efficiency using Eq.(\ref{down}) for different optical depths from $1$
to $300$. \ The maximum is found by varying five parameters mentioned above and the conversion efficiency reaches $100\%$
asymptotically when the optical depth becomes larger. \ In the strong parametric coupling regime as we discussed in the previous
section, $\eta_{\text{d}} \simeq\sin^{2}[\operatorname{Im}(\kappa _{s}L)]$ and it has a maximum when
$\operatorname{Im}(\kappa_{s}L)=\frac{\pi}{2},$ see Fig.\ref{coefficient}.\ Since $\operatorname{Im}(\kappa_{s}L)$ is
proportional to optical depth and inversely proportional to the Rabi frequencies of the driving lasers, an order of magnitude
estimate of the optical depth necessary for near unit conversion efficiency is opd$\simeq\frac{\pi}{2}\Omega_{a,b}/\gamma_{03} >>1$.

\begin{figure}[t]
\begin{center}
\includegraphics[width=5.5cm,height=8cm,angle=-90]{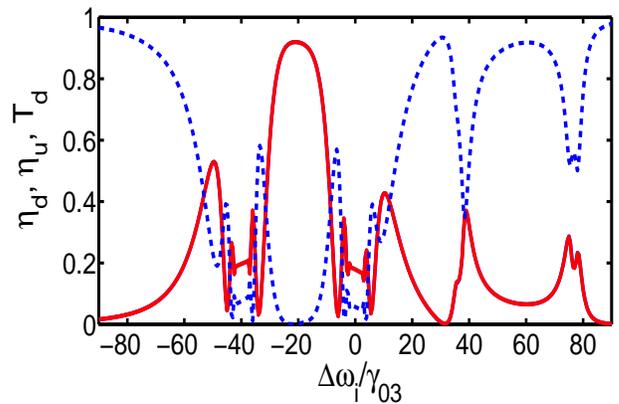}
\caption{(Color online) Conversion efficiency $\eta_{\text{d}}$, $\eta_{\text{u}}$ and transmission $T_{\text{d}}$ vs $\Delta\omega_{i}$ for opd=150. \ $\eta_{\text{d}}$ and $\eta_{\text{u}}$ are indistinguishable and shown in solid red line and $T_{\text{d}}$ is in dashed blue line. \ High transmission efficiency corresponds to low conversion efficiency indicating the approximate conservation condition within each parametric coupling window. The maximum conversion efficiency is found in the left window at around $\Delta\omega_{i}=-20\gamma_{03}$ and other relevant parameters are the same as in Fig.\ref{coefficient}.}%
\label{trans}
\end{center}
\end{figure}

The behavior of the cross-coupling coefficient $\Im(\kappa_{s}L)$ as a function of idler detuning indicates where large
conversion is to be found, as a comparison with Fig.\ref{trans} shows. \ The maximum efficiency of about $0.92$ is located in the
left parametric coupling window at the
 intersection of $\Im(\kappa_{s}L)$ and $\frac{\pi}{2}$. \ Inside the windows the trade-off between conversion and transmission
  is clear. \ In the region where absorption is large, on the sides of the window (especially for the left window), the efficiency
 and the transmission are both low although the valley in conversion efficiency corresponds to a peak in transmission as expected in
  parametric coupling. The transmission approaches unity when the incident idler field is far off-resonance.  
  
  We note that the symmetry $(\Delta_{1},\Delta_{b},\Delta\omega_{i})\rightarrow -(\Delta_{1},\Delta_{b},\Delta\omega_{i})$ gives degenerate optimal conversion conditions.

\section{Pulse conversion: solution of the Maxwell-Bloch equations}

The effect of finite-duration input probe pulses, which are often employed in practice, can be assessed by numerically solving the Maxwell-Bloch equations for the coupled atoms-fields system. The characteristic scales of time and length are given by the Arecchi-Courtens time $T_{c}$ and $L_{c} = cT_{c}$, respectively, which are inversely proportional to the square root of the opd. The cooperative electric field is the product of the atomic number and the idler electric field per photon, i.e., $E_{c} = \sqrt{\rho\hbar\omega_{i}/ (2\epsilon_{0})}.$

Scaling the space, time, and electric field amplitude accordingly, indicated by tildes, the light propagation equation becomes
\begin{equation}
\frac{d}{d\tilde{z}}\tilde{E}_{s}^{+}=i\tilde{\sigma}_{12}\frac{|g_{s}|^{2}}{|g_{i}|^{2}}\text{, \ }
\frac{d}{d\tilde{z}}\tilde{E}_{i}^{+}=i\tilde{\sigma}_{03}.
\end{equation}
A consistent scaling can also be applied to the atomic dynamical equations presented in the Appendix.

\ The Maxwell-Bloch equations were integrated with a semi-implicit finite difference method \cite{semi}. The mid-point integration
method is stable and has high accuracy without sacrificing memory for finer grids \cite{numerical}. \ The algorithm has been
tested by comparing with the parametric equations' solutions in appropriate limits, and these solutions are recovered when
fine enough grids are employed.

To illustrate the influence of finite pump pulse duration, we compute the down conversion efficiency
\begin{equation}
\eta_{\text{d}}=\frac{\int| {E}_{s}^{+}(z=L,t)|^{2} dt}{\int| {E}_{i}%
^{+}(z=0,t)|^{2} dt}.
\end{equation}

\begin{figure}[t]
\begin{center}
\includegraphics[width=7.5cm,height=8.5cm,angle=-90]{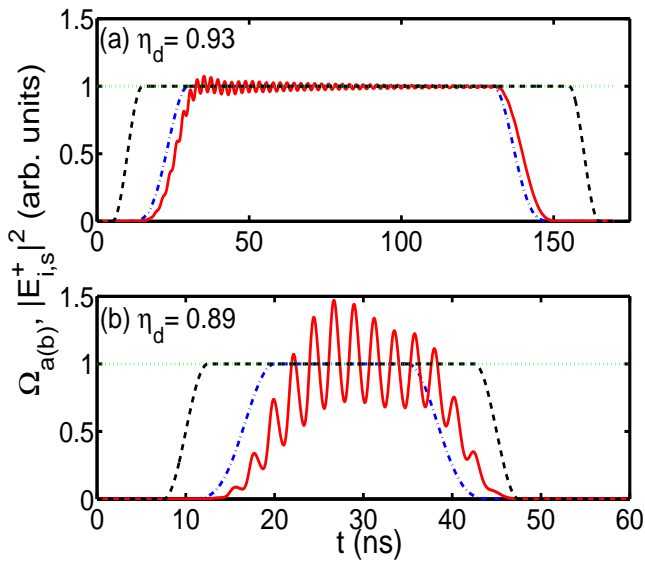}
\caption{(Color online) Time-varying pump fields of Rabi frequencies $\Omega_{a,b}(t)$ and down-converted signal intensity [$|E_{s}^+(t,z=L)|^2$] from an input idler pulse [$|E_{i}^+(t,z=0)|^2$]. Pump-b (dotted green) is a continuous wave and pump-a (dashed black) is a square pulse long enough to enclose input idler with pulse duration (a) 100 ns and (b) 15 ns (dashed-dot blue). Output signal intensity (solid red) at the end of atomic ensemble $z=L$ is oscillatory due to the pump fields. The square pulse in rising region ($t_r-\frac{t_s}{2}<t<t_r+\frac{t_s}{2}$) has the form of $\frac{1}{2}[1+\text{sin}(\frac{\pi(t-t_r)}{t_s})]$ that in (a) $(t_r,t_s)=(10,10)$ns for pump-a and $(t_r,t_s)=(20,20)$ns for input idler; (b) $(t_r,t_s)=(10,5)$ns for pump-a and $(t_r,t_s)=(15,10)$ns for input idler where $t_r$ is the rising time indicating the center of rising period $t_s$. Note that the falling region of square pulse is symmetric to the rising one.}
\label{pulse}
\end{center}
\end{figure}

In Fig. \ref{pulse}, we show the computed values of $\eta_{\text{d}}$ for two different input idler pulse durations. \ We fix the opd$=150$ and use the near optimum parameters ($\Omega_{a}$, $\Omega_{b}$, $\Delta_{1}$, $\Delta_{b}$, $\Delta\omega_{i}$) $=(33$, $20$, $39$, $2$, $-21)\gamma_{03}$ determined from the coupled parametric equations. The temporal shape of the pump laser intensities are also shown. \ Pump laser b is taken to be continuous wave,
while pump a is a square pulse with duration large enough to completely overlap the input idler pulse. \ To compare with the steady state solutions,
we choose the Rabi frequency of idler as $0.1\gamma_{03}$ which is small compare to those of the pumps. \ We find that the conversion efficiency is
reduced for shorter idler pulse inputs.\ \ A $100$ ns idler pulse is long enough that it has \textit{\textbf{a }}almost the same maximum conversion efficiency of \ $0.92$
as in Fig.\ref{opd} for opd=150. \ While for the shorter idler pulse of $15$ ns, the signal develops significant temporal modulation and this reduces the conversion efficiency, although it is still quite appreciable. The modulation frequency is at the generalized Rabi frequency of
  pump-a $\sqrt{\Delta_{1}^{2}+4\Omega_{a}^{2}}$. We note the characteristic time and space scales of the calculations are $T_{c}=0.086$ ns
   and $L_{c}=26$ mm for a moderate atomic density $\rho=1.7\times10^{11}$cm$^{-3}$ and $L=6$ mm.

The grid size for dimensionless time $\Delta\tilde{t}=0.5$ and space $\Delta\tilde{z}=0.001$ were chosen for both $100$ and $15$ ns idler pulse
durations and the convergence is reached with an estimated relative error less than $1\% $.

\section{Conclusion and discussion}

We have studied light frequency conversion in an atomic ensemble with a diamond configuration of atomic levels such as $^{87}$Rb.
The motivation stems from the need to efficiently convert light resonant with ground state transitions (storable in the sense of
quantum memories) to and from the telecom wavelength band for low-loss quantum network communication. The optically thick atomic
sample is driven by two strong co-propagating pump fields, and a probe idler or signal field depending on whether we consider
down- or up-conversion. \ Parametric equations for the probe fields are derived and used to compute conversion efficiencies. By
performing a global search we find conditions of pump Rabi frequencies, detunings and signal/idler input frequency to maximize
the conversion efficiency as a function of optical depth of the ensemble. Only in the limit of very large optical depth does the
maximum efficiency approach the ideal strong coupling result \cite{gogyan}. Under conditions routinely obtained in cold,
non-degenerate rubidium gas, with opd $\simeq 100-200$, optimal conversion efficiencies of the order $80$ to $90 \%$ are
predicted. Numerical solution of the Maxwell-Bloch equations confirms the solution of the parametric equations in the limit of
long pulse duration, and indicates that for shorter pulses, pump pulse induced modulation may reduce the conversion efficiency.

\begin{acknowledgments}*
We acknowledge support from NSF and we thank Alex Kuzmich, Alex Radnaev and Stewart Jenkins for helpful discussions.
\end{acknowledgments}

\appendix
\section{Maxwell-Bloch and parametric coupling equations}

\ To derive the coupled Maxwell-Bloch equations it is convenient to employ a quantized description of the electromagnetic field \cite{quantization}
and use Heisenberg-Langevin equation methods, and then invoke a standard semiclassical factorization assumption. The propagation
length $L$ is discretized into $2M+1$ elements. \ The positive frequency component of the electric field operator is given by
$\hat{E}^{+}(z)=\sum_{n=-M}^{M}\sqrt{\frac{\hbar\omega_{s,n}}{2\epsilon_{0}V}}e^{i(k_{s}+k_{n})z}%
\hat{c}_{n}$ where $[\hat{c}_{n},\hat{c}_{n^{\prime}}^{\dag}]=\delta_{nn^{\prime}}$, $k_{n}=\frac{2\pi n}{L}~,$
$\omega_{s,n}=\omega_{s}+k_{n}c~,$ $n=-M,...,M$ and $\omega_{s}=k_{s}c~$is the central
frequency. \ Define the local boson operators $\hat{a}_{l}=\frac{1}%
{\sqrt{2M+1}}\sum_{n=-M}^{M}\hat{c}_{n}e^{ik_{n}z_{l}}$ where $[\hat{a}_{l},\hat{a}_{l^{\prime}}^{\dag}]=\delta_{ll^{\prime}}$.
Similar definitions hold for the signal, s, and idler field, $i $, which carry an additional index in the following.\

The Hamiltonian for the interacting system depicted in Fig.\ref{four} is given by, (we ignore the interactions responsible for
atomic spontaneous emission for the moment)
\begin{equation}
\hat{H}=\hat{H}_{0}+\hat{H}_{I},%
\end{equation}
where
\begin{align}
&  \hat{H}_{0}\nonumber\\
&  =\sum_{i=1}^{3}\sum_{l=-M}^{M}\hbar\omega_{i}\hat{\sigma}_{ii}^{l}%
+\hbar\omega_{s}\sum_{l=-M}^{M}\hat{a}_{s,l}^{\dag}\hat{a}_{s,l}\nonumber\\
&  +\hbar\sum_{l,l^{\prime}}\omega_{ll^{\prime}}\hat{a}_{s,l}^{\dag}\hat
{a}_{s,l^{\prime}}+\hbar\omega_{i}\sum_{l=-M}^{M}\hat{a}_{i,l}^{\dag}\hat
{a}_{i,l}+\hbar\sum_{l,l^{\prime}}\omega_{ll^{\prime}}\hat{a}_{i,l}^{\dag}%
\hat{a}_{i,l^{\prime}}%
\end{align}
and
\begin{align}
&  \hat{H}_{I}\nonumber\\
&  =-\hbar\sum_{l=-M}^{M}\Big\{\Omega_{a}(t)\hat{\sigma}_{01}^{l\dagger
}e^{ik_{a}z_{l}-i\omega_{a}t}+\Omega_{b}(t)\hat{\sigma}_{32}^{l\dagger
}e^{-ik_{b}z_{l}-i\omega_{b}t}\nonumber\\
&  +g_{s}\sqrt{2M+1}\hat{\sigma}_{12}^{l\dagger}\hat{a}_{s,l}e^{-ik_{s}z_{l}%
}+g_{i}\sqrt{2M+1}\hat{\sigma}_{03}^{l\dagger}\hat{a}_{i,l}e^{ik_{i}z_{l}%
}\nonumber\\
&  +h.c.\Big\},
\end{align}
where $\hat{\sigma}_{mn}^{l}\equiv\sum_{\mu}^{N_{z}}\hat{\sigma}_{mn}^{\mu}\Big|_{z_{\mu}=z_{l}}$, the Rabi frequency $\Omega_{a,(b)}(t)=f_{a,(b)}(t)d_{10,(23)}\mathcal{E}(k_{a,(b)})/(2\hbar)$ is half
the standard definition and $f_{a,(b)}$ is a slowly varying temporal profile without spatial dependence (ensemble size much less
than pulse length). \ The dipole matrix element $d_{mn}\equiv$ $\langle m|\hat{d}|n\rangle$, coupling strength $g_{s,(i)}\equiv
d_{21,(30)}\mathcal{E}(k_{s,(i)})/\hbar,~\mathcal{E}(k)=\sqrt{\hbar\omega/2\epsilon_{0}V}$ and $z_{p}=pL/(2M+1),$ $p=-M,...,M$. \
The matrix $\omega_{ll^{\prime}} \equiv \sum_{{n=-M}^{M}} k_n e^{ik_{n}(z_{l}-z_{l^{\prime}})}/(2M+1)$ accounts for field
propagation by coupling the local mode operators.

\ The dynamical equations including dissipation due to spontaneous emission may be treated by standard Langevin-Heisenberg
equation methods \cite{QO}, and we define $\gamma_{ij}$ as the natural
 transition rate from $|j\rangle \rightarrow | i \rangle.$ Since we are interested in a semiclassical description,
 we replace the field operators by c-numbers in the Langevin equations, and drop the zero-mean Langevin noise sources. All atomic spin
 operators are also replaced by their expectation
 values. Finally, in the co-moving frame coordinates $z$ and $\tau=t-z/c$ the atomic equations are
\begin{widetext}%

\begin{align}
\frac{d}{d\tau}\tilde{\sigma}_{01}  &  =(i\Delta_{1}-\frac{\gamma_{01}}%
{2})\tilde{\sigma}_{01}+i\Omega_{a}(\tilde{\sigma}_{00}-\tilde{\sigma}%
_{11})+ig_{s}^{\ast}\tilde{\sigma}_{02}E_{s}^{-}-ig_{i}\tilde{\sigma}%
_{13}^{\dag}E_{i}^{+}\nonumber\\
\frac{d}{d\tau}\tilde{\sigma}_{12}  &  =(i\Delta\omega_{s}-\frac{\gamma
_{01}+\gamma_{2}}{2})\tilde{\sigma}_{12}-i\Omega_{a}^{\ast}\tilde{\sigma}%
_{02}+ig_{s}(\tilde{\sigma}_{11}-\tilde{\sigma}_{22})E_{s}^{+}+iP^{\ast}%
\Omega_{b}\tilde{\sigma}_{13}\nonumber\\
\frac{d}{d\tau}\tilde{\sigma}_{02}  &  =(i\Delta_{2}-\frac{\gamma_{2}}%
{2})\tilde{\sigma}_{02}-i\tilde{\sigma}_{12}\Omega_{a}+ig_{s}\tilde{\sigma
}_{01}E_{s}^{+}+iP^{\ast}\tilde{\sigma}_{03}\Omega_{b}-iP^{\ast}g_{i}%
\tilde{\sigma}_{32}E_{i}^{+}\nonumber\\
\frac{d}{d\tau}\tilde{\sigma}_{11}  &  =-\gamma_{01}\tilde{\sigma}_{11}%
+\gamma_{12}\tilde{\sigma}_{22}+i\Omega_{a}\tilde{\sigma}_{01}^{\dag}%
-i\Omega_{a}^{\ast}\tilde{\sigma}_{01}-ig_{s}\tilde{\sigma}_{12}^{\dag}%
E_{s}^{+}+ig_{s}^{\ast}\tilde{\sigma}_{12}E_{s}^{-}%
\nonumber\\
\frac{d}{d\tau}\tilde{\sigma}_{22}  &  =-\gamma_{2}\tilde{\sigma}_{22}%
+ig_{s}\tilde{\sigma}_{12}^{\dag}E_{s}^{+}-ig_{s}^{\ast}\tilde{\sigma}%
_{12}E_{s}^{-}+i\Omega_{b}\tilde{\sigma}_{32}^{\dag}-i\Omega_{b}^{\ast}%
\tilde{\sigma}_{32}\nonumber\\
\frac{d}{d\tau}\tilde{\sigma}_{33}  &  =-\gamma_{03}\tilde{\sigma}_{33}%
+\gamma_{32}\tilde{\sigma}_{22}-i\Omega_{b}\tilde{\sigma}_{32}^{\dag}%
+i\Omega_{b}^{\ast}\tilde{\sigma}_{32}+ig_{i}\tilde{\sigma}_{03}^{\dag}%
E_{i}^{+}-ig_{i}^{\ast}\tilde{\sigma}_{03}E_{i}^{-}%
\nonumber\\
\frac{d}{d\tau}\tilde{\sigma}_{13}  &  =(i\Delta\omega_{i}-i\Delta_{1}%
-\frac{\gamma_{01}+\gamma_{03}}{2})\tilde{\sigma}_{13}-i\Omega_{a}^{\ast
}\tilde{\sigma}_{03}-iPg_{s}\tilde{\sigma}_{32}^{\dag}E_{s}^{+}+iP\Omega
_{b}^{\ast}\tilde{\sigma}_{12}+ig_{i}\tilde{\sigma}_{01}^{\dag}E_{i}%
^{+}\nonumber\\
\frac{d}{d\tau}\tilde{\sigma}_{03}  &  =(i\Delta\omega_{i}-\frac{\gamma_{03}%
}{2})\tilde{\sigma}_{03}-i\Omega_{a}\tilde{\sigma}_{13}+iP\Omega_{b}^{\ast
}\tilde{\sigma}_{02}+ig_{i}(\tilde{\sigma}_{00}-\tilde{\sigma}_{33})E_{i}%
^{+}\nonumber\\
\frac{d}{d\tau}\tilde{\sigma}_{32}^{\dag}  &  =(-i\Delta_{b}-\frac{\gamma
_{03}+\gamma_{2}}{2})\tilde{\sigma}_{32}^{\dag}-iP^{\ast}g_{s}^{\ast}%
\tilde{\sigma}_{13}E_{s}^{-}+i\Omega_{b}^{\ast}(\tilde{\sigma}_{22}%
-\tilde{\sigma}_{33})+iP^{\ast}g_{i}\tilde{\sigma}_{02}^{\dag}E_{i}%
^{+}%
\end{align}
\end{widetext}%
where\ $\gamma_{2}=\gamma_{12}+\gamma_{32},$ $P\equiv e^{i\Delta kz-i\Delta\omega t},$ the four-wave mixing mismatch wavevector
$\Delta k=k_{a}-k_{s}+k_{b}-k_{i},$ the frequency mismatch
$\Delta\omega=\omega_{a}+\omega_{s}-\omega_{b}-\omega_{i}=\Delta_{1}-\Delta_{b}+\Delta\omega _{s}-\Delta\omega_{i}$\ and various
detunings are defined as $\Delta\omega_{i}=\omega_{i}-\omega_{3}$, $\Delta\omega_{s}=\omega_{s}-\omega_{12},$
$\Delta_{1}=\omega_{a}-\omega_{1},$ $\Delta_{2}=\omega_{a}+\omega_{s}-\omega_{2}=\Delta_{1}+\Delta\omega_{s}$ ,
$\Delta_{b}=\omega_{b}-\omega_{23}$.
The slow-varying atomic operators are defined%
\begin{align*}
\tilde{\sigma}_{01}  &  \equiv\frac{1}{N_{z}}\hat{\sigma}_{01}^{l}e^{-ik_{a}%
z_{l}+i\omega_{a}t},\tilde{\sigma}_{12}\equiv\frac{1}{N_{z}}\hat{\sigma}_{12}%
^{l}e^{ik_{s}z_{l}+i\omega_{s}t},\\
\tilde{\sigma}_{02}  &  \equiv\frac{1}{N_{z}}\hat{\sigma}_{02}^{l}e^{-ik_{a}%
z_{l}+ik_{s}z_{l}+i\omega_{s}t+i\omega_{a}t},\tilde{\sigma}_{11}\equiv\frac
{1}{N_{z}}\hat{\sigma}_{11}^{l},\\
\tilde{\sigma}_{03}  &  \equiv\frac{1}{N_{z}}\hat{\sigma}_{03}^{l}e^{-ik_{i}%
z_{l}+i\omega_{i}t},\tilde{\sigma}_{22}\equiv\frac{1}{N_{z}}\hat{\sigma
}_{22}^{l},
\end{align*}%
\begin{align*}
\tilde{\sigma}_{32}^{\dag}  &  \equiv\frac{1}{N_{z}}\hat{\sigma}_{32}^{l\dag
}e^{-i\omega_{b}t}e^{-ik_{b}z_{l}},\tilde{\sigma}_{33}\equiv\frac{1}{N_{z}%
}\hat{\sigma}_{33}^{l},\\
\tilde{\sigma}_{13}  &  \equiv\frac{1}{N_{z}}\hat{\sigma}_{13}^{l}e^{-i(\omega
_{a}- \omega_{i})t+i(k_{a}-ik_{i})z_{l}}%
\end{align*}
where $N_{z}(2M+1)=N$ .

The field equations are%
\begin{align}
\frac{d}{dz}E_{s}^{+}  &  =\frac{iNg_{s}^{\ast}}{c}\tilde{\sigma}%
_{12}\\
\frac{d}{dz}E_{i}^{+}  &  =\frac{iNg_{i}^{\ast}}{c}\tilde{\sigma}%
_{03}%
\end{align}
where the field operators are defined as%
\begin{align}
E_{s}^{-}(z,t)  &  \equiv\sqrt{2M+1}\hat{a}_{s,l}^{\dag}e^{-i\omega_{s}%
t},\text{ }\nonumber\\
E_{i}^{+}(z,t)  &  \equiv\sqrt{2M+1}\hat{a}_{i,l}e^{i\omega_{i}t}%
\end{align}

For energy and momentum conservation ($P=1$), and in the\ weak field limit, we solve the atomic equations in steady state after
linearizing with respect to the probe fields
\begin{align}
T_{01}\tilde{\sigma}_{01}  &  =i\Omega_{a}(1-2\tilde{\sigma}_{11}%
-\tilde{\sigma}_{22}-\tilde{\sigma}_{33})\nonumber\\
T_{32}^{\ast}\tilde{\sigma}_{32}^{\dag}  &  =i\Omega_{b}^{\ast}(\tilde{\sigma
}_{22}-\tilde{\sigma}_{33})\nonumber\\
T_{02}\tilde{\sigma}_{02}  &  =-i\Omega_{a}\tilde{\sigma}_{12}+ig_{s}%
\tilde{\sigma}_{01}E_{s}^{+}+i\Omega_{b}\tilde{\sigma}_{03}-ig_{i}%
\tilde{\sigma}_{32}E_{i}^{+}\nonumber\\
T_{13}\tilde{\sigma}_{13}  &  =-i\Omega_{a}^{\ast}\tilde{\sigma}_{03}%
-ig_{s}\tilde{\sigma}_{32}^{\dag}E_{s}^{+}+i\Omega_{b}^{\ast}\tilde{\sigma
}_{12}+ig_{i}\tilde{\sigma}_{01}^{\dag}E_{i}^{+}\nonumber\\
T_{12}\tilde{\sigma}_{12}  &  =-i\Omega_{a}^{\ast}\tilde{\sigma}_{02}%
+ig_{s}(\tilde{\sigma}_{11}-\tilde{\sigma}_{22})E_{s}^{+}+i\tilde{\sigma}%
_{13}\Omega_{b}\nonumber\\
T_{03}\tilde{\sigma}_{03}  &  =-i\Omega_{a}\tilde{\sigma}_{13}+i\tilde{\sigma
}_{02}\Omega_{b}^{\ast}+ig_{i}(\tilde{\sigma}_{00}-\tilde{\sigma}_{33}%
)E_{i}^{+} \label{steady}%
\end{align}
where $T_{01}=\frac{\gamma_{01}}{2}-i\Delta_{1},$ $T_{32}^{\ast}=\frac {\gamma_{03}+\gamma_{2}}{2}+i\Delta_{b},$ $T_{02}=\frac{\gamma_{2}}{2}%
-i\Delta_{2},$ $T_{13}=\frac{\gamma_{01}+\gamma_{03}}{2}+i\Delta_{1}-i\Delta\omega_{i},$ $T_{12}=\frac{\gamma_{01}+\gamma_{2}}{2}-i\Delta
\omega_{s},$ $T_{03}=\frac{\gamma_{03}}{2}-i\Delta\omega_{i}$ and note that $\tilde{\sigma}_{02},$ $\tilde{\sigma}_{13},$ $\tilde{\sigma}_{12},$
$\tilde{\sigma}_{03}$ are expressed in first order of fields and $\tilde{\sigma}_{01},$ $\tilde{\sigma}_{32}^{\dag}$ in zeroth order. \ For
population operators, we solve them in the zeroth order of fields and the nonzero steady states of population and coherence operator are (s denotes
steady state solution)%

\begin{align}
\tilde{\sigma}_{11,s}  &  =\frac{|\Omega_{a}|^{2}}{\Delta_{1}^{2}+\frac
{\gamma_{01}^{2}}{4}+2|\Omega_{a}|^{2}},\text{ }\tilde{\sigma}_{00,s}%
=1-\tilde{\sigma}_{11,s}\nonumber\\
\tilde{\sigma}_{01,s}  &  =\frac{i\Omega_{a}}{\frac{\gamma_{01}}{2}%
-i\Delta_{1}}(1-2\tilde{\sigma}_{11,s})
\end{align}
Substitute the latter results into Eq.(\ref{steady}), and solve for $\tilde{\sigma}_{12}$ and $\tilde{\sigma}_{03}.$ The parametric coupling equations
for the signal and idler fields become%
\begin{align}
\frac{d}{dz}E_{s}^{+}  &  =\beta_{s}E_{s}^{+}+\kappa_{s}E_{i}^{+}\nonumber\\
\frac{d}{dz}E_{i}^{+}  &  =\kappa_{i}E_{s}^{+}+\alpha_{i}E_{i}^{+}
\label{field}%
\end{align}
where%

\begin{align}
\beta_{s}  &  =\frac{-N|g_{s}|^{2}}{cD}[\tilde{\sigma}_{11,s}(T_{03}%
+\frac{|\Omega_{a}|^{2}}{T_{13}}+\frac{|\Omega_{b}|^{2}}{T_{02}})\nonumber\\
&  -\frac{i\Omega_{a}^{\ast}\tilde{\sigma}_{01,s}}{T_{02}}(T_{03}%
+\frac{|\Omega_{a}|^{2}-|\Omega_{b}|^{2}}{T_{13}})]
\end{align}

\begin{align}
\kappa_{s}  &  =\frac{-Ng_{i}g_{s}^{\ast}}{cD}[\tilde{\sigma}_{00,s}%
(\frac{\Omega_{a}^{\ast}\Omega_{b}}{T_{02}}+\frac{\Omega_{a}^{\ast}\Omega_{b}%
}{T_{13}})\nonumber\\
&  +\frac{i\Omega_{b}\tilde{\sigma}_{01,s}^{\dag}}{T_{13}}(T_{03}%
+\frac{|\Omega_{b}|^{2}-|\Omega_{a}|^{2}}{T_{02}})]
\end{align}

\begin{align}
\kappa_{i}  &  =\frac{-Ng_{s}g_{i}^{\ast}}{cD}[\tilde{\sigma}_{11,s}%
(\frac{\Omega_{a}\Omega_{b}^{\ast}}{T_{02}}+\frac{\Omega_{a}\Omega_{b}^{\ast}%
}{T_{13}})\nonumber\\
&  +\frac{i\Omega_{b}^{\ast}\tilde{\sigma}_{01,s}}{T_{02}}(T_{12}%
+\frac{|\Omega_{b}|^{2}-|\Omega_{a}|^{2}}{T_{13}})]
\end{align}

\begin{align}
\alpha_{i}  &  =\frac{-N|g_{i}|^{2}}{cD}[\tilde{\sigma}_{00,s}(T_{12}%
+\frac{|\Omega_{a}|^{2}}{T_{02}}+\frac{|\Omega_{b}|^{2}}{T_{13}})\nonumber\\
&  -\frac{i\Omega_{a}\tilde{\sigma}_{01,s}^{\dag}}{T_{13}}(T_{12}%
+\frac{|\Omega_{a}|^{2}-|\Omega_{b}|^{2}}{T_{02}})]
\end{align}
and
\begin{align}
D  &  \equiv T_{12}T_{03}+T_{12}(\frac{|\Omega_{a}|^{2}}{T_{13}}+\frac
{|\Omega_{b}|^{2}}{T_{02}})+T_{03}(\frac{|\Omega_{a}|^{2}}{T_{02}}%
+\frac{|\Omega_{b}|^{2}}{T_{13}})\nonumber\\
&  +\frac{(|\Omega_{a}|^{2}-|\Omega_{b}|^{2})^{2}}{T_{02}T_{13}}.%
\end{align}


\begin{thebibliography}{99}                                                                                               %

\bibitem {repeater}H.-J. Briegel, W. D\"{u}r, J. I. Cirac, and P. Zoller, Phys. Rev. Lett. 81, 5932 (1998)

\bibitem {eit}S. E. Harris, Phys. Today 50, No. 7, 36 (1997)

\bibitem {Lukin}M. Fleischhauer and M. D. Lukin, Phys. Rev. Lett. 84, 5094 (2000)

\bibitem {polariton}S. D. Jenkins, D. N. Matsukevich, T. Chaneli\`{e}re, A. Kuzmich, and T. A. B. Kennedy, Phys. Rev. A 73, 021803(R) (2006)

\bibitem {qubit}D. N. Matsukevich and A. Kuzmich, Science 306, 663 (2004)

\bibitem {chou} C. W. Chou, S. V. Polyakov, A. Kuzmich, and H. J. Kimble, Phys. Rev. Lett. 92, 213601 (2004)

\bibitem {entangle}D. N. Matsukevich, T. Chaneli\`{e}re, M. Bhattacharya, S.-Y. Lan, S. D. Jenkins, T.A.B. Kennedy, and A. Kuzmich, Phys. Rev. Lett. 95, 040405 (2005)

\bibitem {vuletic}A. T. Black, J. K. Thompson, and V. Vuleti\'{c}, Phys. Rev. Lett., 95, 133601 (2005)

\bibitem {retrieval}T. Chaneli\`{e}re, D. Matsukevich, S. D. Jenkins, S.-Y. Lan, T.A.B. Kennedy, and A. Kuzmich, Nature 438, 833 (2005)

\bibitem {single}D. N. Matsukevich, T. Chaneli\`{e}re, S. D. Jenkins, S.-Y. Lan, T.A.B. Kennedy, and A. Kuzmich, Phys. Rev. Lett. 97, 013601 (2006)

\bibitem {pan}S. Chen, Y.-A. Chen, T. Strassel, Z.-S. Yuan, B. Zhao, J. Schmiedmayer, and J.-W-. Pan, Phys. Rev. Lett. 97, 173004 (2006)

\bibitem {kimble}J. Laurat, H. de Riedmatten, D. Felinto, C.-W. Chou, E. W. Schomburg, and H. J. Kimble, Opt. Exp. 14, 6912 (2006)

\bibitem {ionent}D. L. Moehring, P. Maunz, S. Olmschenk, K. C. Younge, D. N. Matsukevich, L-M. Duan, and C. Monroe, Nature 449, 68 (2007)

\bibitem {ion} D. N. Matsukevich, P. Maunz, D. L. Moehring, S. Olmschenk, and C. Monroe, Phys. Rev. Lett. 100, 150404 (2008)

\bibitem {radaev}A. G. Radnaev, Y. O. Dudin, R. Zhao, H. H. Jen, S. D. Jenkins, A. Kuzmich, and T. A. B. Kennedy, Nature Physics, in press.

\bibitem {ran}R. Zhao, Y. O. Dudin, S. D. Jenkins, C. J. Campbell, D. N. Matsukevich, T. A. B. Kennedy, and A. Kuzmich, Nature Phys. 5, 100 (2009)

\bibitem {telecom}T. Chaneli\`{e}re, D. N. Matsukevich, S. D. Jenkins, T.A.B. Kennedy, M.S. Chapman, and A. Kuzmich, Phys. Rev. Lett. 96, 093604 (2006)

\bibitem {gorshkov}A. V. Gorshkov, A. Andr\'{e}, M. D. Lukin, and A. S. S\o rensen, Phys. Rev. A. 76, 033804 (2007) 

\bibitem {orozco}F. E. Becerra, R. T. Willis, S. L. Rolston, and L. A. Orozco, Phys. Rev. A 78, 013834 (2008)

\bibitem {braje}D. A. Braje, V. Bali\'{c}, S. Goda, G. Y. Yin, and S. E. Harris, Phys. Rev. Lett. 93, 183601 (2004)

\bibitem {couple}M. D. Lukin, A. B. Matsko, M. Fleischhauer, and M. O. Scully, Phys. Rev. Lett. 82, 1847 (1999)

\bibitem {gogyan08}A. Gogyan and Yu. Malakyan, Pyhs. Rev. A 77, 033822 (2008)

\bibitem {gogyan}A. Gogyan, Phys. Rev. A 81, 024304 (2010)

\bibitem {gsgi}O. S. Heavens, J. Opt. Soc. Am., Vol. 51, 1058 (1961)

\bibitem {scale}F. T. Arecchi and E. Courtens, Phys. Rev. A 2, 1730 (1970)

\bibitem {semi}P.D. Drummond, Comp. Phys. Comm. 29, 211 (1983)

\bibitem {numerical}W. H. Press, S.A. Teukolsky, W. T. Vetterling and B. P. Flannery, Numerical Recipes in C, 2nd Edition (Cambridge Univ. Press, 1992)

\bibitem {quantization}P. D. Drummond and S. J. Carter, J. Opt. Soc. Am. B, Vol. 4, 1565 (1987)

\bibitem {QO}M. O. Scully and M. S. Zubairy, Quantum optics (Cambridge Univ. Press, 1997)

\end{thebibliography}
\end{document}